\documentstyle[12pt]{article}
\begin{document}
\newcommand{\beq}{\begin{equation}}
\newcommand{\eeq}{\end{equation}}
\begin{titlepage}
\begin{center}
{\Huge{ITF}}\hfill UUITP $8$/1994\\
{\Huge{Uppsala Universitet}}\hfill hep-th/9412172\\
\rule[.1in]{13.5cm}{.01in} \\
\hfill  \\
\vspace{0.1in}{\Large\bf On the stability \\of  $p$-brane}\\[.4in]
{\large  P.Demkin}
\footnote{On leave from Department of Physics,
Vilnius University, Saul\.{e}tekio al.9, 2054,\\ Vilnius, Lithuania}\\
\bigskip {\it
Institute of Theoretical Physics \\
Uppsala University\\
Box 803, S-75108\\
Uppsala, Sweden\\
paul@rhea.teorfys.uu.se}

\bigskip \bigskip
\end{center}
\begin{abstract}
Stability of some solutions of the equations of motion of  bosonic

$p$-branes in curved and flat spacetimes is stated.

\vspace{0.1in}
\end{abstract}

\end{titlepage}
\newpage
\section{Introduction}
Besides strings as one-dimensional extended relativistic objects, we may
 also consider $p$-branes as $p$-dimensional surfaces moving in
 spacetime. Such a membrane model naturally appears (i) when
 generalizing the known shell-electron model, which has already been
 suggested by Dirac \cite{D,Ha,Ba} ;
 (ii) as a cosmic domain wall in the post-inflationary universe
 \cite{L, La} ; (iii) as an effective model of supergravity \cite{Hu} ;
 and (iv) as, like superstring, a model unifying fundamental
 interactions \cite{Be, B}.

Let us turn to the last point. Unlike the properties of strings, those of
 $p$-branes are much less investigated so far \cite{Du,D2}. The quantum
 properties of supermembrane are known only on the semiclassical level.
 The spectrum continuity of supermembrane has been treated as its
 quantum and even classical instability, as its degenerate turning into an
 infinite string without changing its energy \cite{W,Fu} or else as its total
 instability. Therefore, the term "instability" in this context means
 asymptotical behaviour of $p$-brane solution at $t\rightarrow \infty$.

The existence of stable $p$-brane solution is important to the general theory
 of extended objects. For further development of the theory
 (quantization, perturbation theory and so on) we have to be sure that
 there is at least one example of stable solution of the equation of motion.

This paper shows that the class of stable $p$-brane solutions is not empty.
 For this purpose, we shall consider some of the solutions of $p$-brane
 equations of motion, both new and known ones \cite{Li,H}, in curved
 and flat spacetimes.

It is necessary to agree upon the main term 'stability' in advance.
 There are many kinds of stability that are known in mathematics
 (structural, Poisson, Lagrange, conditional, absolute
 $etc.$). Solutions of equations of motion are their stable
 points respecting their mappings. We shall restrict our
 consideration only to the asymptotical behaviour of the $p$-brane
 solution at $t\rightarrow \infty$ and to Lyapunov and asymptotical
stabilities.
 The stable point $x_0$ of the mapping $A$ is Lyapunov stable (and,
 respectively, asymptotically stable), if $\forall \varepsilon >
 0$, so that if $|x-x_0|< \delta $, then $|A^nx-A^nx_0|<
 \varepsilon$ for all $0< n< \infty $ (correspondingly, $A^nx-A^nx_0
 \rightarrow 0$, as $n \rightarrow \infty $).

\section{Bosonic membrane in curved spacetime}
We may separate the bosonic part of the membrane in $D=11$ by
 extinguishing fermionic degrees of freedom of the supermembrane.
 The action for the supermembrane becomes the action for a purely
 bosonic membrane in a curved spacetime:
\begin{eqnarray}
S=-\frac{T}{2}\int d^3\xi \sqrt {h}[h^{ij}\partial _{i}x^M
\partial _jx^Ng_{MN}(x)-\nonumber\\
-\frac{1}{3}\varepsilon ^{ijk} \partial _ix^M
\partial _jx^N \partial _kx^PB_{MNP}(x)-1],
\end{eqnarray}
where $h_{ij}=\partial _{i}x^M\partial _jx^N
g_{MN},
\quad h=|deth_{ij}|, \quad g_{MN}(x)$ is a metric of curved spacetime,
 $B_{MNP}(x)$ is an antisymmetric tensor field of rank 3, which couples with
membrane
via a Wess-Zumino term \cite{Be,B,D2}.

In this case, the equations of motion for membrane turn into
\begin{eqnarray}
\partial _i(\sqrt{h}h^{ij}\partial_jx^Ng_{MN})
-\frac{1}{2}\sqrt{h}h^{ij}
\partial _ix^N
\partial _jx^P\partial _Ng_{MP}-\nonumber\\
-\frac{1}{6}\varepsilon ^{ijh} \partial _ix^N \partial _jx^P
\partial _kx^QF_{MNPQ}=0,\qquad \quad
\end{eqnarray}
where $F_{MNPQ}=4\partial _{[M}B_{NPQ]}$ is the tension of

 potential $B_{MNP}$.

Let us consider the solution for the bosonic membrane in $D=11$ with a
special spherical spacetime symmetry metric $g_{MN}(x)$
\begin{eqnarray}
ds^{2}=-e^{2a}dt^{2}+e^{2b}[dr^{2}+r^{2}d\Omega ^{2}]+e^{2c}[dy+g\cos
 \theta d\varphi -qdt]^{2}+\nonumber\\+(dx^{5})^{2}+...+(dx^{10})^{2}
\qquad
 \qquad \qquad \qquad \;
 \end{eqnarray}
in the form
\begin{equation}
\xi ^0=t,\;  \xi ^1=r, \;  \xi ^2=y;\quad X^3=\theta, \; X^4=\varphi ,\;X^{N}
\neq f(\xi ), \: N=5,...,10.
\end{equation}

Using the explicit expression for metric (3) and choosing  coordinates
 (4) we obtain the equations of motion for the parameters $a, b, c, q$:
\begin{eqnarray}
(a^{''}+\frac{2}{r}a^{'})e^{-2b}-\frac{1}{2}{q^{'}}
^2e^{-4(a+b)}=0,\\
({a^{'}}^2+{b^{'}}^2+\frac{1}{r}b^{'}+a^{'}b^{'})e^{-2b}-
\frac{1}{4}{q^{'}}
^2e^{-4(a+b)}=0,\\
(b^{''}+\frac{2}{r}b^{'})e^{-2b}+\frac{g^2}{2r^4}
e^{-2a-6b}=0,\\
\left[a^{''}+b^{''}+\frac{2}{r}(a^{'}+b^{'})\right]
e^{-2b}-
\frac{1}{2}{q^{'}}^2e^{-4(a+b)}+\frac{g^2}{2r^4}
e^{-2a-6b}=0,\\
q^{''}+(\frac{2}{r}-4a^{'}-2b{'})q^{'}e^{-2a-3b}=0
\end{eqnarray}
with the extra condition $a+b+c=const$.

The solutions to these equations are:

1. The stable monopole solution found by D.J.Gross and M.J.Perry
 \cite{Gr}:
\begin{eqnarray}exp(2b)=exp(-2c)=1+\bar{g}/r, \quad   a=0,
 \quad   q=0, \quad   g^{2}={\bar{g}}^{2}.
\end{eqnarray}

2. Its electrically charged analog
\begin{eqnarray}
exp(-2a)=exp(2c)=1+\bar{A}/r,\quad b=0,\nonumber\\
q=-(A/r)(1+\bar{A}/r)^{-1},\quad g=0,\quad A^2=\bar{A}^2.
\end{eqnarray}

3. Its dyon analog
\begin{eqnarray}
exp(-2a)=exp(2b)=(1+\bar{g}/r)^2,\quad c=0,\nonumber\\
q=-(g/r)(1+\bar{g}/r)^{-1},\quad g^2=2\bar{g}^2.
\end{eqnarray}

When considering a purely bosonic membrane, in the
 supersymmetric action we can extinguish fermionic degrees of
 freedom, $i.e.$ spacetime gravitino $\psi _{M}(X)$ and fermionic
 coordinates $\theta (\xi )$, or else let them equal zero. In this case
 we obtain a bosonic sector of the supermembrane in a curved
 spacetime. We have just considered such a way and obtained
 stable monopol-like solutions. The Nambu-Goto action in a
 curved spacetime is another way to consider a purely bosonic
 membrane. We may start from the Nambu-Goto action
\begin{equation}
S_D=-T\int d^{p+1}\xi \sqrt{|det\partial _\alpha X^\mu
 \partial _\beta X^
\nu g_{\mu \nu}|},
\end{equation}
where $g_{\mu \nu}=g_{\mu \nu}(X)$.

The equations of motion derived from this action have the following
 form:
\begin{equation}
\partial _\alpha \left(\sqrt{\gamma} \gamma^{\alpha \beta}
 \partial _\beta x^\nu g_{\mu \nu}\right)=\frac{1}{2}
\sqrt{\gamma} \gamma^{\alpha \beta}
\partial _\alpha X^\nu \partial _\beta X^\lambda
\frac{\partial g_{\nu \lambda}}{\partial X^\mu},
\end{equation}
where $\gamma_{\alpha \beta}$ is completely defined by the induced
 metric
\begin{equation}
\gamma_{\alpha \beta}=\partial_\alpha X^\mu \partial _\beta
X^\nu g_{\mu \nu}.
\end{equation}

Let us consider the simplest spherically simmetric non-flat metric,
 $i.e.$ the Schwarzschild solution of the metric
\begin{eqnarray}
g_{\mu \nu}=\left(\begin{array}{cccc} 1-q & 0 & 0 & 0\\0 & {-(1-q)^{-1}}
 & 0 & 0 \\ 0 & 0 & -r^2 & 0 \\ 0 & 0 & 0 & -r^2\sin ^2 \theta
  \end{array}\right)
\end{eqnarray}
where $q=2GM/r$, $G$ is Newton's constant, and $M$ is the total
 gravitational mass of the membrane.

For the $p=2$ membrane in the spherical coordinate system in $D=4$,
 $X^\mu =\left( \tau , r(\tau ), \theta , \phi  \right),$ the equation
of motion (14) becomes
\begin{equation}
(1-q)r\ddot{r}-2(1-q)\dot{r}^2+\frac{1}{4}q\dot{r}^3+\frac{1}{4}
q(1-q)^2\dot{r}+2(1-q)^3=0.
\end{equation}

At $q=1$ it gets into the black-hole region. In the case when $q\leq 1$
 (close to unity), we may decompose the solution $r(\tau )$  near the
 Schwarzschild radius $r_0=2GM (r=r_0+\rho)$. Then the equations of
motion (17) turn into
\begin{eqnarray}
\left[(r_0+\rho )\rho \ddot{\rho}/r_0+\frac{1}{4}\dot{\rho}^3-
2\rho \dot{\rho}^2/r_0\right](1+\rho /r_0)^{-1}+\\\left(\frac{1}{4}
\rho ^2 \dot{\rho}/r_0 ^2+2\rho ^3/r^3_0\right)(1+\rho/r_0)^{-3}=0
\nonumber.
\end{eqnarray}

We may consider this equation accounting for the terms that make the
 greatest contribution into the evolution of $\rho=\rho (\tau )$:
\begin{equation}\rho \ddot{\rho}+2\rho^3/r^3_0=0.
\end{equation}

The solution of this equation is
\begin{equation}
\rho(\tau)=\left[\rho^{-\frac{1}{2}}(0)+\frac{1}{\sqrt{3}r^\frac{3}{2}_0}
(\tau -\tau _0)\right]^{-2}.
\end{equation}

The velocity of such membrane motion is
\begin{equation}
\dot{\rho}(\tau)=-\frac{2}{\sqrt{3}}(\rho /r_0)^{\frac{3}{2}},
\end{equation}
$i.e.$ we obtain an asymptotical fall of the membrane on the
 Schwarzschild sphere during the infinite time.

\section{Bosonic $p$-brane in flat spacetime}
The equations of motion for bosonic relativistic $p$-brane in a
 flat spacetime are
\begin{equation}
\partial_{\alpha}(\sqrt{h}h^{\alpha \beta}\partial_{\beta}
X^{\mu })=0.
\end{equation}
These equations and the constraint conditions
\begin{equation}
P^{\mu}_{\tau}X_{\mu ;i}=0,  \qquad  P^{2}+T^{2}deth_{ij}=0,
\end{equation}
where $P^{\mu}_{\tau}=\delta \cal L / \delta $\.{X}$^{\mu},
 \quad 1\leq i,j\leq p$, $\; \;   $and the border conditions
 for open dimensions
\begin{equation}
\frac{\partial h(\sigma _{i}^{in})}{\partial X^{\mu}_{;i}}=
\frac{\partial h(\sigma _{i}^{f})}{\partial X^{\mu}_{;i}}=0,
\qquad
\sigma _{i}^{\in}\left[ \sigma _{i}^{in},\sigma _{i}^{f}  \right]
\end{equation}
may be obtained from the action
\begin{equation}
S = -T\int d^{p+1}\zeta \sqrt {|det(\partial_{\alpha}X^{\mu}\partial_{\beta}
X_{\mu})|}.
\end{equation}

Equations of motion (22) are equivalent to the ordinary Laplace
 operator $\triangle $ on a non-compact manifold $\Sigma ^{p+1}$,
 $i.e.$  on the worldvolume of the $p$-brane. We know that the
 Laplace operator in the Euclidean spacetime is a stable operator
 under small perturbations near the  solution $X^{\mu }_{0}$ as
 a linear elliptic operator. However, the worldvolume
 metric $h_{\alpha \beta }$ is a pseudo-Euclidean metric, and the
 stability in this case is yet unclear.

In order to have the physical picture of surface $\Sigma ^{p+1}$,
 we may choose $X^{0}=\tau $ and obtain
\begin{equation}
X^{\mu }=(\tau ,X^{m}(\tau ,\sigma _{1},...,\sigma _{p})),
\qquad m=1,...,D-1.
\end{equation}
Then the metric $h_{\alpha \beta }$ is
\begin{equation}
h_{\alpha \beta }=\left(\begin{array}{cc}
{\dot X^{2}-1}&
{\dot{X}^{m}\partial _{a}X_{m}}\\
{\dot{X}^{m}\partial _{a}X_{m}}&
{\partial _{a}X^{m}\partial_{b}X_{m}}\end{array}\right).
\end{equation}

Without any restriction we may use the reparametrization invariance
 of the action and suggest $h_{0 \beta }=\dot{X}^{m}\partial _{b}
 X_{m}=0.$
Then the metric tensor is
\begin{equation}
h_{\alpha \beta }=\left(\begin{array}{cc}
{\dot X^{2}-1}&
{0}\\
{0}&
{\partial _{a}X^{m}\partial_{b}X_{m}}\end{array}\right).
\end{equation}

The equations of motion become
\begin{eqnarray}
\ddot X^{m}+\frac {1}{2}\partial _{a}\dot X^{2}h^{ab}\partial _{b}
X^{m}+(\dot X^{2}-1)\triangle X^{m}=0,\\
\partial _{\tau }\left(\frac{\sqrt{h}}{\sqrt{1-\dot X^{2}}}\right)=0,
\qquad \qquad \dot X^{2}\leq 1,
\end{eqnarray}
where $\triangle =\frac{1}{\sqrt{h}}\partial _{a}(\sqrt{h}h^{ab}
\partial _{b})$ is the Laplace operator on the  space part of the
 metric tensor $h_{ab}$. Now that we have defined the Laplace operator

 on an appropriate Euclidean subspace, we may treat

 these equations of motion as a general dynamic system.

Let us introduce new variables $Y^{m}=\dot X^{m}$ and consider small
 variations $\delta X^{m}=\xi ^{m}$ and $\delta Y^{m}=\eta ^{m}$
 respecting the solution $X^{m}_{0},Y^{m}_{0}$. For these variations
 we have
\begin{eqnarray}
\dot \xi ^{m}=\eta ^{m},\qquad \qquad \qquad \\
\dot \eta ^{m}=\frac{\partial Q^{m}(X_{0},Y_{0})}{\partial X^{n}}\xi
 ^{n}+\frac{\partial Q^{m}(X_{0},Y_{0})}{\partial Y^{n}}\eta ^{n},
\end{eqnarray}
where $Q^{m}=\frac{1}{2} \partial _{a}Y^{2}h^{ab}\partial _{b}X^{m}$.

At $D=1$, the behaviour of the dynamical system is known very well. In
 this case, the solutions of the
 characteristical equation $\lambda ^{2}-Q'_{X}(X_{0},Y_{0})=0$ are
 purely imaginary, and this means that the point $(X_{0},Y_{0})$ is
 either the centre or the focus. A necessary and sufficient condition
 for the existence of the centre is that the initial dynamical system
 must have the time-independent, real and holomorphous integral
 $F(X,Y)=C$ in the neighbourhood of the point $(X_{0},Y_{0})$. In
 this case, the point $(X_{0},Y_{0})$ corresponds to the minimum of
 the potential energy of the system. It seems likely that in an
 analytical system with $D$ degrees of freedom, the solution of the
 equations of motion, which is not a minimum point, is unstable. It
 is a pity that this has not been proved for $D\geq 2$.

Let us consider a special solution to the equations of motion for a
 closed surface that is a mixture of pulsation and rigid rotation:
\begin{equation}
X^{m}(\tau ,\sigma _{1},...,\sigma _{p})=x(\tau )(\cos \varphi (\tau
  )n^{k},\sin \varphi (\tau )n^{l},0,...,0),
\end{equation}
where $n^{k},n^{l}=(n_{1},...,n_{d})$ is a $d$-dimensional unit vector
 describing the embedding of a $p$-dimensional closed surface in
 $S^{d-1}$ and $d\leq (D-1)/2$.

In this case, equations (30) correspond to
\begin{equation}
\dot x^{2}+x^{2}\dot \varphi ^{2}+\left(\frac{x}{C}\right)^{2p}=1,
\end{equation}
where $C$ is a constant with the length dimension  allowing to bring
 the equation for $x$ into the form
\begin{equation}
\ddot x-x^{2p-2}/C^{2p}\triangle x=0.
\end{equation}
$\triangle =\frac{1}{\sqrt{g}}\partial _{a}(\sqrt{g}g^{ab}
\partial _{b})$
 is the covariant Laplacian on the $p$-surface swept out by $\vec{n}$,
 and $g_{ab}=\partial _{a}n\partial _{b}n$ is the corresponding
 metric.

Let $L=x^{2}\dot \varphi =const$ corresponding to the conservation of
 the angular momentum. Then the equations of motion turn into
\begin{eqnarray}
(\triangle +p)n^{k}=0,\qquad \quad \\
\ddot x-L^{2}/x^{3}+px^{2p-1}/C^{2p}=0.
\end{eqnarray}

The first equation is Laplace operator in Euclidean space
 $R^{d}$. If the metric of the surface is known, this is an
 ordinary linear operator on a compact manifold, and it is stable.
 In the general case, this is the  equation of minimal surface in
 the Euclidean space. From the general theory of minimal surfaces
 we know that solutions to this equation do exist, and some of the
 solutions $n_{0}^{k}$ are stable under small variations near
 $n_{0}^{k}$ \cite {Dao}. The second equation is an ordinary
 differential equation equivalent to
\begin{eqnarray}
y=\dot x,\qquad \qquad \qquad \qquad \\
\dot y=f(x), \quad f(x)=L^{2}/x^{3}-px^{2p-1}/C^{2p}, \quad
 f'(x_{0})< 0.
\end{eqnarray}

The characteristical equation $\lambda ^{2}-f'(x_{0})=0$ has purely
 imaginary solutions. The energy conservation law in the form
\begin{equation}
\left(\frac{x}{C}\right)^{2p}=1-\dot x^{2}-\frac{L^{2}}{x^{2}}\geq 0
\end{equation} following from (34) is the holomorphous integral
 $F(X,Y)=C$ in the neighbourhood of the point $(X_{0},Y_{0})$,
 $i.e.$ the necessary and sufficient condition for the centre to
 exist. Its  Poincar\'{e} index equals to unity.

Equation of motion (37) for the radial part $x(\tau)$ depends on two
 parameters, but in fact we may neglect one of them.  Condition (40)
 is valid for both every time moment $\tau $ and initial conditions.
 From (40), it follows that
\begin{equation}
{\tt y}^2=1-k{\tt x}^{2p}-\frac{1}{\tt x^2},
\end{equation}
where ${\tt x}=\frac{x}{L}$ is the new dimensionless variable, ${\tt y}
=\dot {\tt x}, \quad k=\left(\frac{L}{x_0}\right)^{2p}\times \
 \times\left(1-\dot x_0^2-\frac{L^2}{x_0^2}\right)$. The phase diagrams
 of this equation are represented in Fig.1.

\section{Discussion}
In the general case, the last word concerning stability belongs to
 the behaviour of the second variation of the action in the solution
 point. However, when the solution is known, we may check its
 stability "by hand" considering small perturbations, as it has been
 done above.

Thus, we have stable solutions for bosonic p-brane in curved and flat
 spacetime. All of them obey the constraints (23). They prove the existence of
 stable solutions for further development of the theory of relativistic
extended
 objects.

Now one may get the impression that author claims all known $p$-branes
 to be stable and compact. This is not so.
 For example, the elegant solution of equations (29)-(30),
 found by Kikkawa-Yamasaki
 \cite{Ki} and generalized by Hoppe \cite{H},
\begin{equation}
X(\tau ,\vec {\sigma })=(f_1(\vec {\sigma }) \vec{n}_1(\tau ), ...,
 f_p(\vec {\tau }) \vec {n}_p(\tau ), 0, ...,0),
\end{equation}
where $\vec {n}_r(\tau )=(cos \omega _r \tau , sin \omega_r \tau )$
 and $f_p(\vec{\sigma })$ are arbitrary functions having
 to satisfy $\Sigma^p_{r-1} \omega ^2_r f^2_r(\vec{\sigma })=1$, is
 non-compact. In this case, indeed, it is possible to change the
 space configuration of the $p$-brane without changing its energy,
 because, according to the last condition, $f_r(\vec{\sigma })
 \rightarrow \infty $ is possible at $\omega _r \rightarrow 0$.

It is very useful to compare the properties of the stability
 of solution (33) in Euclidean spacetime. In this
 case, equations of motion (29), (30) turn into equations
\begin{eqnarray}
\ddot X^{m}+\frac {1}{2}\partial _{a}\dot X^{2}h^{ab}\partial _{b}
X^{m}+(\dot X^{2}+1)\triangle X^{m}=0,\\
\partial _{\tau }\left(\frac{\sqrt{h}}{\sqrt{1+\dot X^{2}}}\right)=0,
\qquad \qquad \dot X^{2}\leq 1,
\end{eqnarray}

Substituting expression $X(\xi )$ (33) turns them into
\begin{eqnarray}
(\triangle +p)n^{k}=0,\qquad \quad \\
\ddot x-L^{2}/x^{3}-px^{2p-1}/C^{2p}=0.
\end{eqnarray}

The first equation is the same as (36) in Minkowski
 spacetime, \- $i.e.$ the equation of the minimal surface for the space
 part of the worldvolume metric, but equation (46) is of another
 character. The new variables ${\tt x}, {\tt y}$, like in the case
 of Minkowsi spacetime, turn equation (41) into
\begin{equation}
{\tt y}^2=-1+k{\tt x}^{2p}-\frac{1}{\tt x^2}.
\end{equation}
The phase diagrams of this equation are represented in Fig.2. These
 diagrams correspond to unstable solutions of the equation of
 motion with the potential
\begin{equation}
f(x)=\frac{L^2}{x^3}+p\frac{x^{2p-1}}{C^{2p}},
\end{equation}
where ${\tt x}=\frac{x}{L}$ is a new dimensionless variable, ${\tt
 y}=\dot {\tt x}, \quad k=\left(\frac{L}{x_0}\right)^{2p}\times \
 \times\left(1+\dot x_0^2+\frac{L^2}{x_0^2}\right)$ ,
 and all the solutions are non-compact for any $p\in {\bf Z_{+}}$ and $k>
 0$.

Stability is only one among many other classical
 requirements imposed on $p$-branes. For example, border
 conditions (24) are very hard restrictions for open $p$-brane.
 In this connection we can mention that the solution
 of the equation of motion for cylindrical membrane found in \cite{Co}
 does not obey the
 border condition. Any physically appropriate solution must have
 no singularity.
 The equations of motion for spherical bosonic membrane (17) at
 $q=0$ in the form
\begin{equation}
r\ddot{r}-(D-2)r\dot{r}^2+(D-2)=0,\qquad
   \dot{r}(r)=\frac{1}{r_0^{D-2}}
\left[r_0^{2(D-2)}-r^{2(D-2)}\right ]
^\frac{1}{2}
\end{equation}
have a peculiarity at $r=0$: they collapse to zero.

As is known, stable classical solutions may be destroyed by
 quantum fluctuations. Thus, the next important question is the
 observance of the stability requirement on the quantum level.

Essential nonlinearity of the $p$-brane equations of motion
 makes it difficult to investigate their quantum stability. The
 solutions that are known to us have been considered only in the
 semiclassical approach \cite{D2,PD,D1}, whereas the question of stability
 will be fully answered only if the quantum stability is proved
 beyond the frames of the perturbation theory.

\section{Acknowledgements}
Author wants to express his gratitude to the Swedish Institute for
 Grant 304/01 GH/MLH, which gave him the opportunity to enjoy the
 warm hospitality of Prof.Antti Niemi, Doc.Staffan Yngve and all the
 members of Institute of Theoretical Physics, Uppsala
 University.

\end{document}